# Nanoscale ferroelectricity in pseudo-cubic sol-gel derived barium titanate - bismuth ferrite (BaTiO$_3$- BiFeO$_3$) solid solutions

A. Pakalniškis[1] A. Lukowiak[2], G. Niaura[3], P. Głuchowski[2,4], D. V. Karpinsky[5,6], D. O. Alikin[8,9], A.S. Abramov[8], A. Zhaludkevich[5], M. Silibin[6,7], A.L. Kholkin[8,9], R. Skaudžius[1], W. Strek[2], A. Kareiva[1,*]

[1]Institute of Chemistry, Vilnius University, Naugarduko 24, LT-03225 Vilnius, Lithuania
[2]Institute of Low Temperature and Structure Research, Polish Academy of Sciences, Okolna 2, PL-50422 Wroclaw, Poland
[3]Institute of Chemical Physics, Faculty of Physics, Vilnius University, Sauletekio Ave. 9, LT-10222, Vilnius Lithuania
[4]Nanoceramics Spolka Akcyjna, Okolna 2, PL-50422 Wroclaw, Poland
[5]Scientific-Practical Materials Research Centre of NAS of Belarus, 220072 Minsk, Belarus
[6]National Research University of Electronic Technology "MIET", 124498 Moscow, Russia
[7]Institute for Bionic Technologies and Engineering, I.M. Sechenov First Moscow State Medical University, Moscow 119991, Russia
[8]School of Natural Sciences and Mathematics, Ural Federal University, Ekaterinburg, Russia
[9]Department of Physics & CICECO – Aveiro Institute of Materials, University of Aveiro, Aveiro, Portugal;

**Abstract**

Single phase barium titanate–bismuth ferrite ((1-x)BaTiO$_3$-(x)BiFeO$_3$, BTO-BFO) solid solutions were prepared using citric acid and ethylene glycol assisted sol-gel synthesis method. Depending on the dopant content the samples are characterized by tetragonal, tetragonal-pseudocubic, pseudocubic and rhombohedral structure as confirmed by Raman spectroscopy and XRD measurements. An increase of the BFO content leads to a reduction in the cell parameters accompanied by a decrease in polar distortion of the unit cell wherein an average particle size increases from 60 up to 350 nm. Non zero piezoresponse was observed in the compounds with pseudocubic structure while no polar distortion was detected in their crystal structure using X-ray diffraction method. The origin of the observed non-negligible piezoresponse was discussed assuming a coexistence of nanoscale polar and non-polar phases attributed to the solid solutions with high BFO content. A coexistence of the nanoscale regions having polar and non-polar character is considered as a key factor to increase macroscopic piezoresponse in the related compounds due to increased mobility of the domain walls and phase boundaries.

## 1. Introduction

At room temperature bismuth ferrite is a multiferroic having a rhombohedral perovskite structure described by R3c space group [1]. While having both its ferroelectric Curie temperature Tc ~ 1100 K and antiferromagnetic Néel temperature $T_N$ ~ 640 K it has attracted a lot of attention [2,3]. As a multiferroic, it can be used in magnetic sensors [4], energy harvesting devices [5] or memory devices [6]. As a piezoelectric material, it is a potential substitute for currently most used $PbZr_xTi_{1-x}O_3$ due to enormously high polarization being measured in the form of thin films [7]. While being a more ecological material since it contains no lead, it is additionally good candidate for high temperature piezoelectric applications due to its high Curie temperature [8,9].

The most critical problem of $BiFeO_3$ is large leakage current significantly reduce applications and partially determined by poor phase stability [3]. Synthesis of single-phase bismuth ferrite is a difficult procedure because none-perovskite secondary phases of $Bi_{25}FeO_{40}$ and $Bi_2Fe_4O_9$ are formed during the fabrication process [10,11]. To avoid the problem of secondary phase formation many approaches were undertaken like using different synthesis methods such as hydrothermal [12], sol-gel [13], mechanochemical method [14] and more. It is also reported that pure bismuth ferrite can be obtained by using extremely pure oxides as precursors with purity over 99.999 % [15]. The third way of stabilization for $BiFeO_3$ structure was to make solid solutions with other perovskite material. While the latter method of stabilization is useful it also affects properties of the original bismuth ferrite phase [16]. On the other hand, $BaTiO_3$ is one of the most well-known ferroelectric materials with low leakage and is easy to be sintered by a liquid chemistry route [17].

These two materials seem to be very promising for formation of solid solution due to the enhancement of polarization, stabilization of the structure and improving overall piezoelectric performance of the ceramics.

In this work, we report on citric acid and ethylene glycol assisted sol-gel synthesis method for the preparation of single phase $(1-x)BaTiO_3-(x)BiFeO_3$ (BTO-BFO) solid solutions. X-ray diffraction analysis and Raman spectroscopy were used for the determination of phase purity. The cell parameters were calculated using the results of Rietveld refinement based on the X-ray diffraction data. The surface morphology of sol-gel derived BTO-BFO solid solutions and piezoelectric properties are also investigated and discussed.

## 2. Experimental

Analytical grade chemicals of $Bi(NO_3)_3 \cdot 5H_2O$, $C_{12}H_{28}O_4Ti$, $Fe(NO_3)_3 \cdot 9H_2O$, $Ba(CH_3COO)_2$, ethylene glycol and citric acid were used as starting materials. For a typical synthesis of 1 g final product the following procedure has been carried out. Firstly, citric acid was dissolved in 20 ml of distilled water at a molar ratio of 3:1 to the final cation amount at 80 °C. Secondly, titanium isopropoxide was added to the above solution. Next, barium acetate, iron nitrate, bismuth nitrate were dissolved in the same solution. Finally, when all materials have been dissolved, the 4 ml of ethylene glycol was added to the present solution. Then the solution was stirred for 1.5 h and evaporated at 200 °C. Obtained gel was then dried at 220 °C overnight. Then xerogel was ground in an agate mortar and heated in a furnace at 650 °C for 5 h with a heating rate of 1 °C/min.

X-ray diffraction (XRD) analysis was performed using Rigaku MiniFlex diffractometer on a glass sample holder. Measurements were performed using Cu K$\alpha$ $\lambda$ = 1.541874 Å radiation measuring from 10° to 70° while moving 10°/min.

Raman spectra were recorded using inVia Raman (Renishaw, United Kingdom) spectrometer equipped with thermoelectrically cooled (–70 °C) CCD camera and microscope. Raman spectra were excited with 532 nm beam. Parameters of the bands were determined by fitting the experimental spectra with Gaussian-Lorentzian shape components using GRAMS/A1 8.0 (Thermo Scientific, USA) software.

Scanning electron microscopy (SEM) images were taken for the morphology characterization with Hitachi SU-70 SEM.

Piezoresponse force microscopy measurements was used to characterize local piezoelectric properties. Experiments have been carried out using MFP-3D commercial scanning probe microscope (Oxford Instruments, UK). The measurements were performed with 17 N/m spring constant, 10 nm tip radius commercial HA_HR Scansens tips with $W_2C$ coating under ac voltage with the amplitude Vac = 5 V and frequency f = 20 kHz. Calibration of the probe tip displacements and cantilever displacements in PFM measurements were made by the following methods described in [18]. Amplitude of the out-of-plane PFM response obtained from quasi-static calibrations was divided by shape factor and amplitude of AC voltage excitation in order to evaluate effective $d_{33}$ coefficient. Corresponding correction of R·CosΘ piezoresponse signal was done before by phase shift maximizing in-phase R·CosΘ signal and minimize out-of-phase R·SinΘ signal [19]. Other signals for all images can be found separately in supplementary materials.

## 3. Results and discussion

The $BaTiO_3$ (BTO) and $BiFeO_3$ (BFO) solid solutions were prepared using different molar ratio of components (BTO:BFO = 1:9, 1:4, 3:7, 2:3, 1:1, 3:2, 7:3, 4:1 and 9:1). The XRD patterns of nine different solid solutions are given in Fig. 1. as a contour map. The blue colour indicates the lowest intensity (background) meanwhile the red colour represents the most intensive points (the peaks). The black columns designate the reference XRD data of BaTiO$_3$ taken from the crystallography open database. According to the PDF (COD 96-150-7757) data the desired products were obtained no matter the ratio of solid solution has been chosen. Nevertheless, a slight shift of the peaks towards lager 2θ values is observed upon increase of BFO content which indicates a decrease in unit cell parameters.

For the deeper analysis of structure development, the Rietveld refinement was employed. The cell parameters calculated by Rietveld analysis are presented in Fig. 2. In general, barium titanate exists in three different structures – cubic (C-phase), tetragonal (T-phase) and trigonal (rhombohedral axes, labelled as R-phase), meanwhile bismuth ferrite belongs to trigonal crystal system with hexagonal or rhombohedral axes. Calculated *a* and *c* parameters for primitive lattice allow to classify the solid solutions by the crystal structure. Note that, during the structure refinement of R-phase *a* and *c* parameters were recalculated in order to obtain the reduced values which are closer to each other and easy to compare.

The following equations were used:

$$a(reduced) = \frac{a(primitive)}{\sqrt{2}} \quad (1)$$

$$c(reduced) = \frac{c(primitive)}{2\sqrt{3}} \quad (2)$$

According to Rietveld analysis data three different blocks of different structures are identified. The phase transitions from tetragonal to cubic and finally to trigonal with rhombohedral axes are observed with increasing amount of BFO.

The phase transitions in the investigated system were also observed by Raman spectroscopy. The results give additional information and confirming the results of X-ray diffraction measurements. Raman spectroscopy is able to provide detailed information on short range structure or local symmetry. Fig. 3 compares Raman spectra of bulk $BaTiO_3$ and $BiFeO_3$. The sharp band near 308 cm$^{-1}$ associated with $B_1$ and $E$ symmetries of longitudinal optical (LO) and transverse optical (TO) phonon modes [$B_1$, $E$(TO+LO)] and high frequency band near 715 cm$^{-1}$ [$A_1$, $E$(LO)] are characteristic for $BaTiO_3$ ferroelectric phase with tetragonal symmetry [20–22]. It should be noted that observed bands corresponds to several phonons because frequencies of the modes are very close [20]. The other dominant bands are relatively broad features located at 257 cm$^{-1}$ [$A_1$(TO)] and 518 cm$^{-1}$ [$A_1$, $E$(TO)]. All observed bands are characteristic for $BaTiO_3$ [20–25]. The intensity of Raman bands of $BiFeO_3$ decreases by a factor of 20 comparing with spectrum of $BaTiO_3$ (Fig. 3). Four sharp characteristic Raman bands of $BiFeO_3$ are visible at 77 cm$^{-1}$ ($E$), 141 cm$^{-1}$ ($A_1$), 174 cm$^{-1}$ ($A_1$), and 220 cm$^{-1}$ ($A_1$) [26]. Theoretical analysis has indicated that Bi atom participates mainly in vibrational modes lower than 167 cm$^{-1}$, while oxygen atoms are involved in vibrational modes higher than 262 cm$^{-1}$ [27]. The broad band near 1256 cm$^{-1}$ involves oxygen atom stretching vibrations. Similar high frequency band is clearly visible in the spectra of lepidocrocite (γ-FeOOH) and maghemite (γ-$Fe_2O_3$) at 1300 and 1360 cm$^{-1}$, respectively [28,29]. It was suggested that relative intensity of these bands depends on the excitation wavelength (resonance enhancement) [28,29]. The intense band near 1310 cm$^{-1}$ was also observed in the spectrum of haematite (α-$Fe_2O_3$) [28,29].

Fig. 4 demonstrates the dependence of Raman spectra on the composition of (1- x)$BaTiO_3$-(x)$BiFeO_3$ solid solution structures. Introduction of 10 % of $BiFeO_3$ results in considerable spectral changes; first of all, the sharp peak near 308 cm$^{-1}$ completely disappears, indicating phase transformation has started.
Anyway, the tetragonal $BaTiO_3$ phase in this new structure is still the dominant phase according to the crystal lattice data obtained by Rietveld analysis. In addition, the peak at 518 cm$^{-1}$ shifts to 511 cm$^{-1}$ and a new low-frequency band near 186 cm$^{-1}$ appears. Such spectral changes are similar to previously observed Fe-doping induced formation of distorted tetragonal/cubic phase BaTiO$_3$ structure [25]. The 186-cm$^{-1}$ peak might be associated with the presence of small amount of TiO$_2$ anatase phase undetectable by XRD measurements and which is usually visible in the low crystalline samples [25]. An increase in intensity and broadening of 724-cm$^{-1}$ band points on the presence of Ba$^{2+}$ defects in the BaTiO$_3$ lattice [25]. Similar Raman bands with progressive decrease in intensity were observed with increasing x part up to 0.3 (Fig. 4). Addition of higher BiFeO$_3$ amount results in changes in the Raman spectrum indicating alterations in the local lattice structure. No clear bands characteristic to BiFeO$_3$ is visible in the low-frequency spectral range; however, the broad feature due to oxygen atom stretching vibrations appears near 1355 cm$^{-1}$ at x = 0.6. This band clearly shifts to lower wavenumbers with increasing content of BiFeO$_3$. Such frequency shift indicates changes in the geometry of oxygen octahedra around the Fe cations. In addition, the new band appears near 681−683 cm$^{-1}$ reaching the highest relative intensity at x = 0.7 of BiFeO$_3$ content. The results again confirm the constructed phase diagram.

SEM micrographs of all samples are given in Fig. 5. The size of the particles was measured using open-source Fiji software by accidentally choosing appropriate particles [30]. It is clearly seen that with increasing the amount of BFO the particle size also increases. The particle size varies from 60 to 120 nm for the samples which SEM images are presented in Fig. 5A, B, C and D. Additionally, the boundaries between the particles vanished. The particles start to gain a more distinct shape with a larger size which in some cases exceeds over 350 nm for the sample with the equal ratio of BTO and BFO (1:1) (Fig. 5E) and for the samples with higher amount of BFO in the solid solutions (Fig. 5F, G H and I). This could be related with the changes of the crystal structure. The change from tetragonal to cubic and finally to trigonal structure causes the formation of particles with bigger size. Note that, independently on the ratio of BTO and BFO in the solid solution the large size distribution of the particles was observed. The semispherically shaped particles have formed when barium titanate is dominating in the solid solution and rectangular particles are predominating in the samples with increasing amount of bismuth ferrite.

Finally, the sol-gel method leads to the formation of slightly agglomerated irregular spherical-rectangular shape particles with rather broad size distribution [31]. The particle size is dependent on the molar ratio of constituents in the BTO-BFO solid solutions. The increase of the particle size is caused by the different melting points of BFO and BTO. It has been previously reported that mixing the higher melting point component, in this case $BaTiO_3$, with another component having a lower melting point, in this case $BiFeO_3$, leads to better crystallinity and improved particle growth [32].

Analysis of the piezoelectric properties of $(0.4)BaTiO_3$-$(0.6)BiFeO_3$, $(0.3)BaTiO_3$-$(0.7)BiFeO_3$ and $(0.2)BaTiO_3$-$(0.8)BiFeO_3$ composition at Bi-rich side was done locally by piezoresponse force microscopy [18]. Piezoresponse was analysed before and after local poling by ±35 V DC voltage. Local switching of the polarization has been done by scanning of rectangular areas with positively and negatively biased tip. Fig 6 demonstrates out-of-plane PFM images before and after local poling for three discussed compositions. In the PFM RCosΘ images, the contrast corresponds to value and sign of the effective $d_{33}$ coefficient. Thereby, "bright" areas represent domains with approximately upward polarization orientation, meanwhile dark contrast areas represent the opposite case (approximately downward polarization). It is clearly seen from these series of images that increase of the $BaTiO_3$ content in solid solution tends to degrade piezoelectric properties of the material. Before poling both $(0.2)BaTiO_3$ –$(0.8)BiFeO_3$ and $(0.3)BaTiO_3$-$(0.7)BiFeO_3$ compositions revealed clusters of polar phase with high effective $d_{33}$ and clusters with piezoresponse close to zero, while (0.4)BaTiO$_3$ –(0.6)BiFeO$_3$ composition didn't show any distinguishable response. Surprisingly, after poling bi-polar contrast can be observed in all three compositions. This is indicative of partial polarization switching across the rectangular area.

It must be noted that small size of the grains can significantly act onto results of PFM measurements due to limitation of PFM spatial resolution. Close to zero piezoresponse inspected before poling can be sourced by effect of averaging the piezoresponse from amount of nanosized domains with different polarization orientation. After poling all disordered polarization states become aligned and thereby can be probed by PFM. However, due to meta-stable structural state of BFO-BTO solid solution we cannot exclude as well that electric field can induce phase transition from non-polar state to polar as such transformation is likely in rare earth doped BFO [33].

Further we analysed in-plane piezoresponse at the smaller scale. In-plane piezoresponse is indicative of piezoelectric activity and excludes most of the known PFM parasitic contributions [34]. The behaviour of in-plane response was similar to out-of-plane (Fig. 7). Comparison of the piezoresponse with topography in (0.2)BaTiO$_3$–(0.8)BiFeO$_3$ composition with largest grains revealed that individual grains before poling consisted of small scale domains, while after poling the polarization become aligned (fig 7a, d). At the same time, we didn't reveal any transformation of the phase without piezoelectric response to piezoelectrically active phase in this composition. Following the decrease of the grain size with increase of the BaTiO$_3$ concentration domains as well became smaller and finally indistinguishable by PFM in (0.4)BaTiO$_3$-(0.6)BiFeO$_3$ composition. After local poling all composition revealed clear bi-polar domain pattern as well as an out-of-plane response.

Fraction of polar phase can be roughly estimated from PFM histograms according approach from [35]. It was found to be around 95 % for (0.2)BaTiO$_3$-(0.8)BiFeO$_3$ composition and decreases down to 18 % for (0.4)BaTiO$_3$-(0.6)BiFeO$_3$ composition. This trend of piezoelectrically active area decrease is followed by median effective d$_{33}$ value reduce (fig 7g-i). This is qualitatively fit to macroscopically observed trend of rhombohedral-pseudocubic structural transformations revealed by XRD measurements. Thus, we postulate that phase macroscopically identified as pseudo-cubic is actually in coexistence of the local nanoscale phases similar to known phase coexistence at morphotropic phase boundary and polymorphic phase boundary in different piezoelectric materials [36]. Further insight into the details of this unusual phase coexistence can be obtained by using methods with enhanced spatial resolution and sensitivity. The XRD method used to describe the phase transformation is not well suitable to characterize nanoscale domains observed by PFM method. While one can observe notable widening of the X-ray diffraction reflections in the region ascribed to the pseudo cubic phase (Fig. 2) which points at a decrease in the average size of the crystallites and support the results obtained by local scale measurements. To conclude, an increasing amount of BaTiO$_3$ leads to degradation of piezoresponse, probably caused by decrease of piezoelectrically active phase amount and a gradual change into a more symmetric pseudo cubic structure induced by BTO. Transformation of the cell to centrosymmetric state extracted from XRD must be followed by cell dipole moment reduction and, consequently, degradation of macroscopic polarization and effective piezoelectric coefficient. We confirmed here this trend by effective d$_{33}$ measurements. Nevertheless, for all measured samples the polarization was switchable, meaning that each solid solution retains ferroelectric properties.

## 4. Conclusions

In conclusion, a systematic study on the structure, morphology and piezoelectric properties of BTO-BFO solid solutions was performed. The Rietveld analysis data has demonstrated that introducing bismuth ferrite into the barium titanate matrix leads to the structural evolution from tetragonal to (pseudo)cubic and finally to trigonal (with rhombohedral axes). All phase structure modifications are concluded in the structure phase diagram. Moreover, increasing amount of BFO in the solid solutions causes not only the structure modifications but it also induces a formation of larger sized, distinctly shape particle which exceed in some cases over 350 nm. Domain structure corresponds to grain size and domains become large towards to Bi rich boundary in the solid solution. Surprisingly, compositions nominally being centrosymmetric exhibit ferroelectricity that was shown to be sourced by nanosized structural states clearly visible after local poling. The explored piezoresponse force measurements have demonstrated that an increasing amount of $BaTiO_3$ leads to degradation of piezoresponse of the solid solution.

**Acknowledgments**

The work has been done in frame of the project TransFerr. This project has received funding from the European Union's Horizon 2020 research and innovation programme under the Marie Sklodowska-Curie grant agreement No. 778070. The scanning probe microscopy study was funded by RFBR (grant No. 19-52-04015) and BRFFR (grant No. F19RM-008). The equipment of the Ural Center for Shared Use "Modern nanotechnology" UrFU was used. Sample structural characterization was funded by RFBR (grant #18-38-20020 mol_a_ved). M.S. also acknowledges Russian academic excellence project "5-100" for Sechenov University. This work was developed within the scope of the project CICECO-Aveiro Institute of Materials, POCI-01-0145-FEDER-007679 (FCT Ref. UID/CTM/50011/2013), financed by national funds through the FCT/MEC and when appropriate co-financed by FEDER under the PT2020 Partnership Agreement.

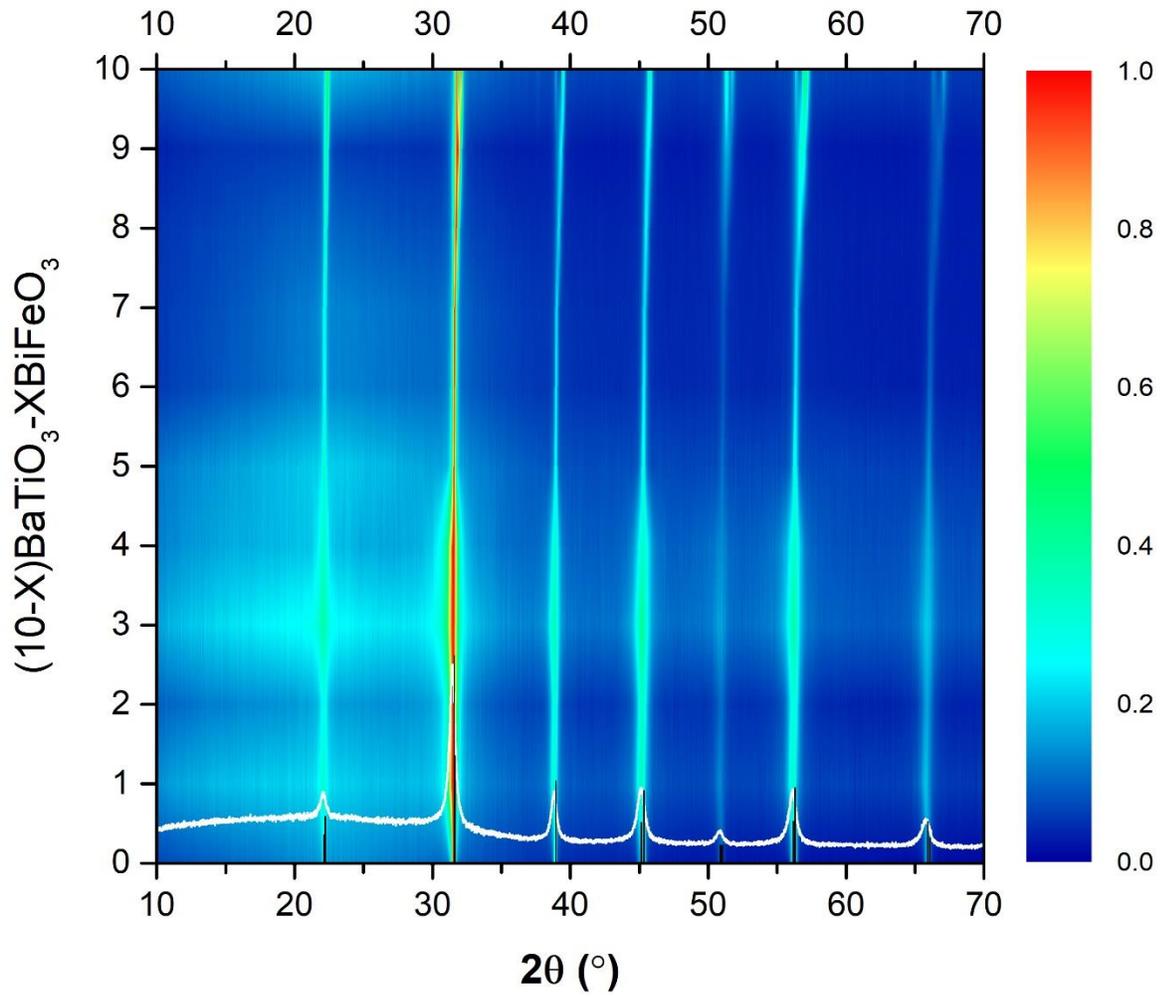

**Fig. 1.** XRD data of (10-x)BaTiO$_3$ – (x)BiFeO$_3$ solid solutions, where 0 < x < 10.

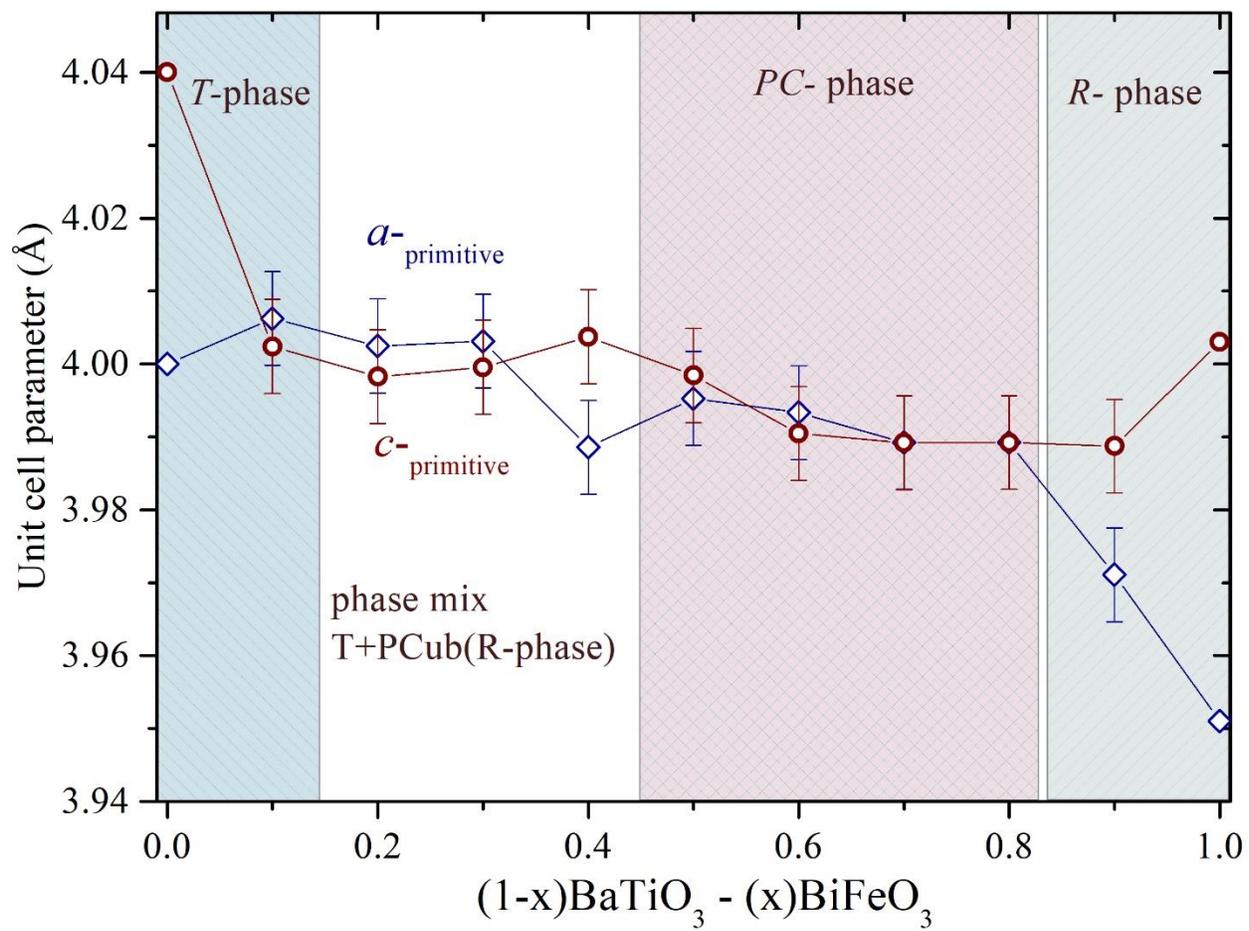

**Fig. 2.** Phase diagram of (1-x)BaTiO$_3$ – (x)BiFeO$_3$ solid solutions.

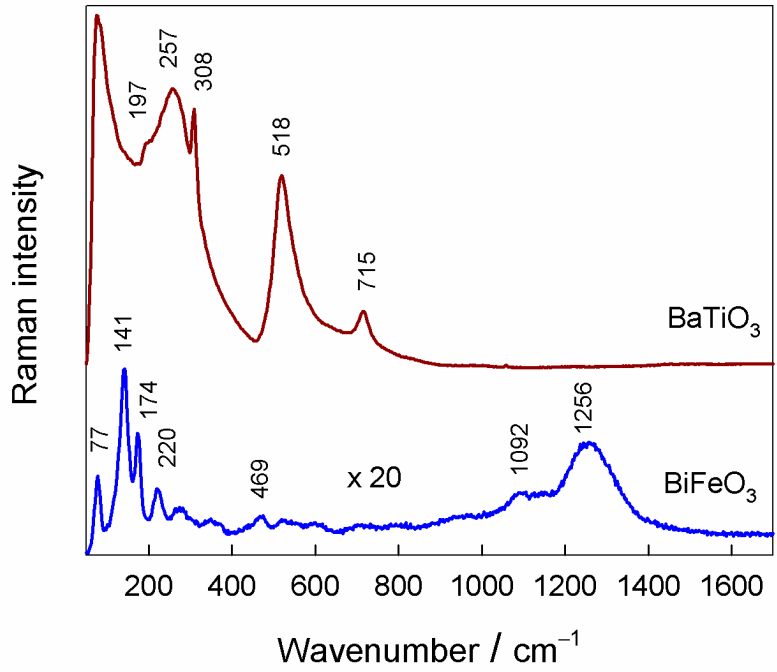

**Fig 3.** Raman spectra of bulk BaTiO$_3$ and BiFeO$_3$ compounds. The excitation wavelength is 532 nm (0.3 mW).

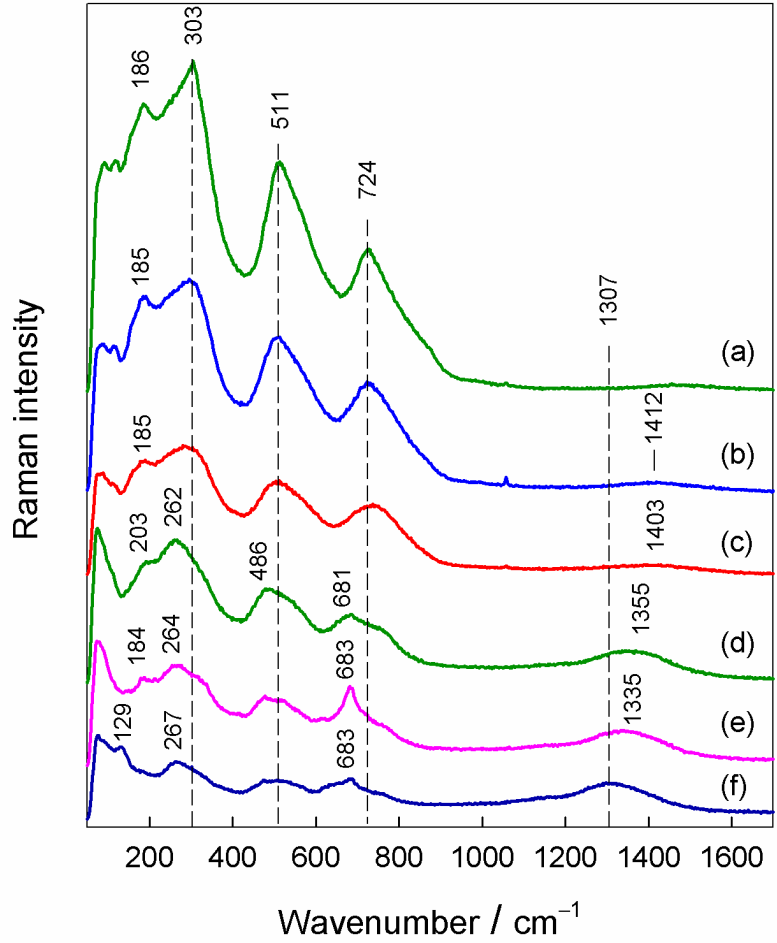

**Fig. 4.** Composition dependent Raman spectra of BTO-BFO solid solutions: (a) (0.9)BaTiO$_3$-(0.1)BiFeO$_3$; (b) (0.8)BaTiO$_3$-(0.2)BiFeO$_3$; (c) (0.7)BaTiO$_3$-(0.3)BiFeO$_3$; (d) (0.4)BaTiO$_3$-(0.6)BiFeO$_3$; (e) (0.3)BaTiO$_3$-(0.7)BiFeO$_3$; (f) (0.2)BaTiO$_3$-(0.8)BiFeO$_3$. The excitation wavelength is 532 nm (0.3 mW).

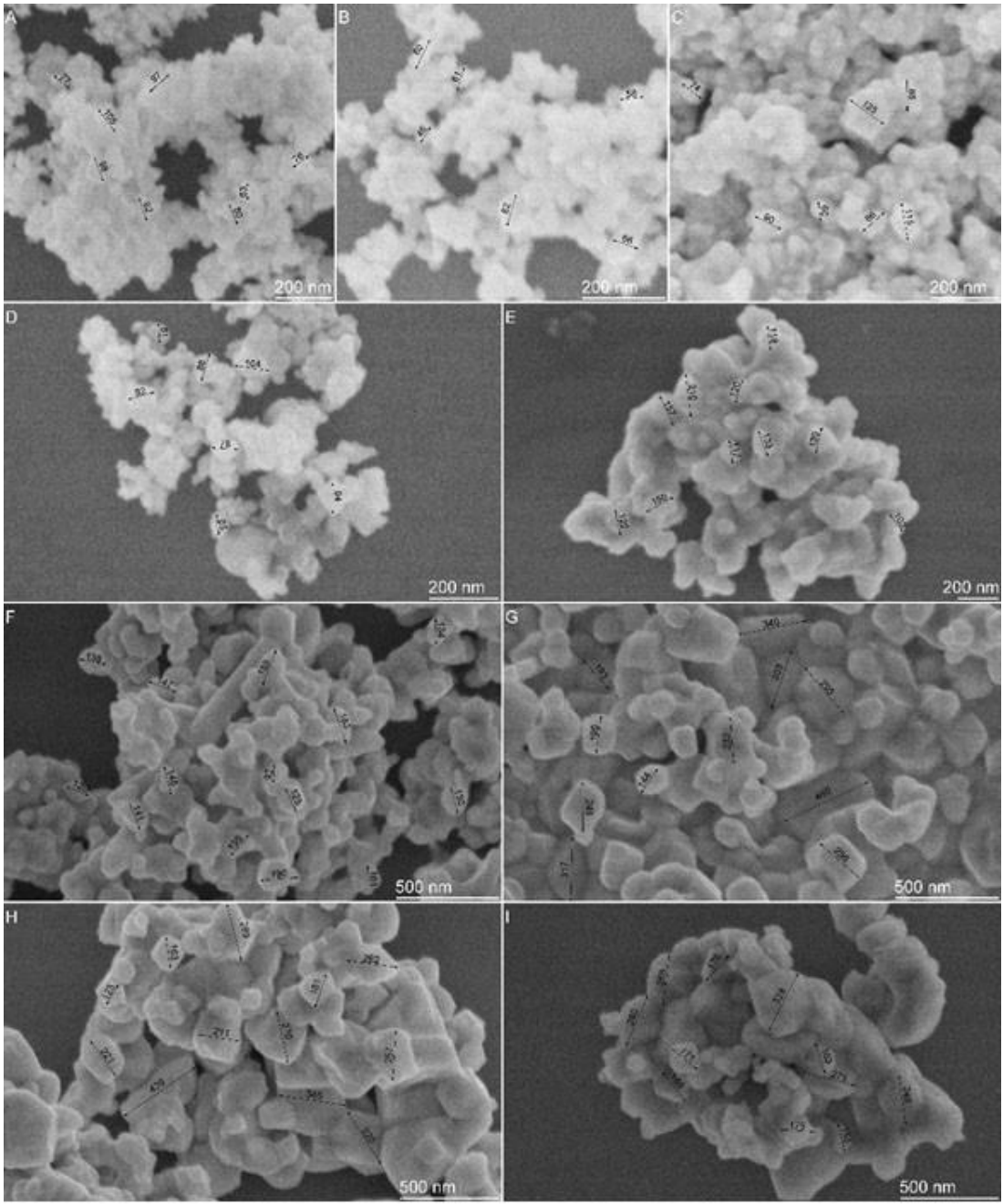

**Fig. 5**. SEM images of BTO-BFO solid solutions: A - (0.9)BaTiO$_3$–(0.1)BiFeO$_3$, B - (0.8)BaTiO$_3$–(0.2)BiFeO$_3$, C - (0.7)BaTiO$_3$–(0.3)BiFeO$_3$, D - (0.6)BaTiO$_3$–(0.4)BiFeO$_3$, E - (0.5)BaTiO$_3$–(0.5)BiFeO$_3$, F - (0.4)BaTiO$_3$–(0.6)BiFeO$_3$, G - (0.3)BaTiO$_3$–(0.7)BiFeO$_3$, H - (0.2)BaTiO$_3$–(0.8)BiFeO$_3$ and J - (0.1)BaTiO$_3$–(0.9)BiFeO$_3$.

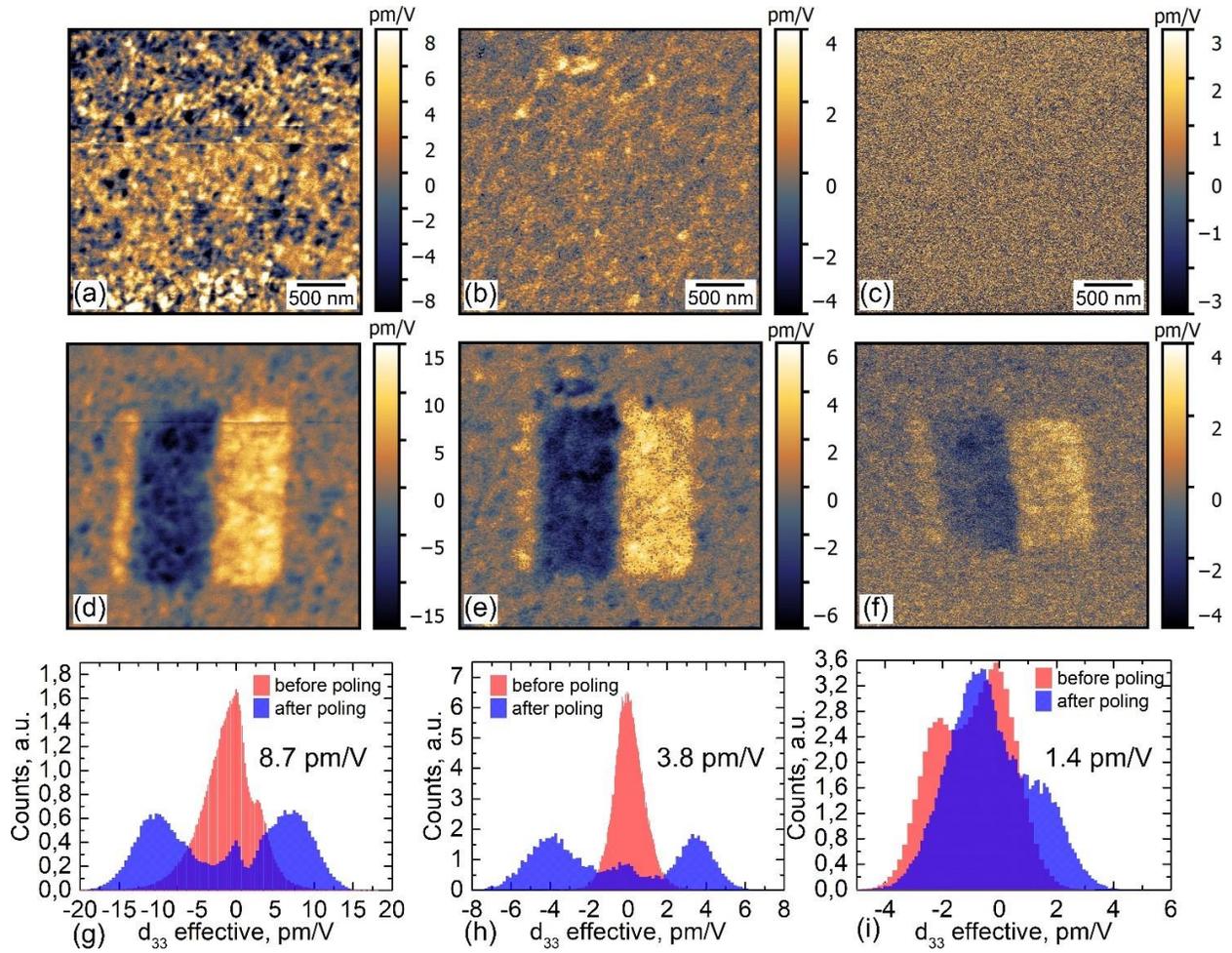

**Fig. 6.** Quantified out-of-plane PFM images. RCosθ piezoresponse signal with meaning of the effective $d_{33}$ coefficient: (a)-(c) before and (d)-(f) after local poling of bi-square area by ±35 V DC voltage (left part is poled negatively, while right part - positively). (g)-(i) Corresponding histograms of effective $d_{33}$ distribution across scan area and inside poled region. (a), (d), (g) $0.8BiFeO_3 - 0.2BaTiO_3$; (b), (e), (h) $0.7BiFeO_3 - 0.3BaTiO_3$; (c), (f), (i) $(0.6)BiFeO_3-(0.4)BaTiO_3$-solid solutions. Calculated median effective $d_{33}$ is displayed at (g)-(i).

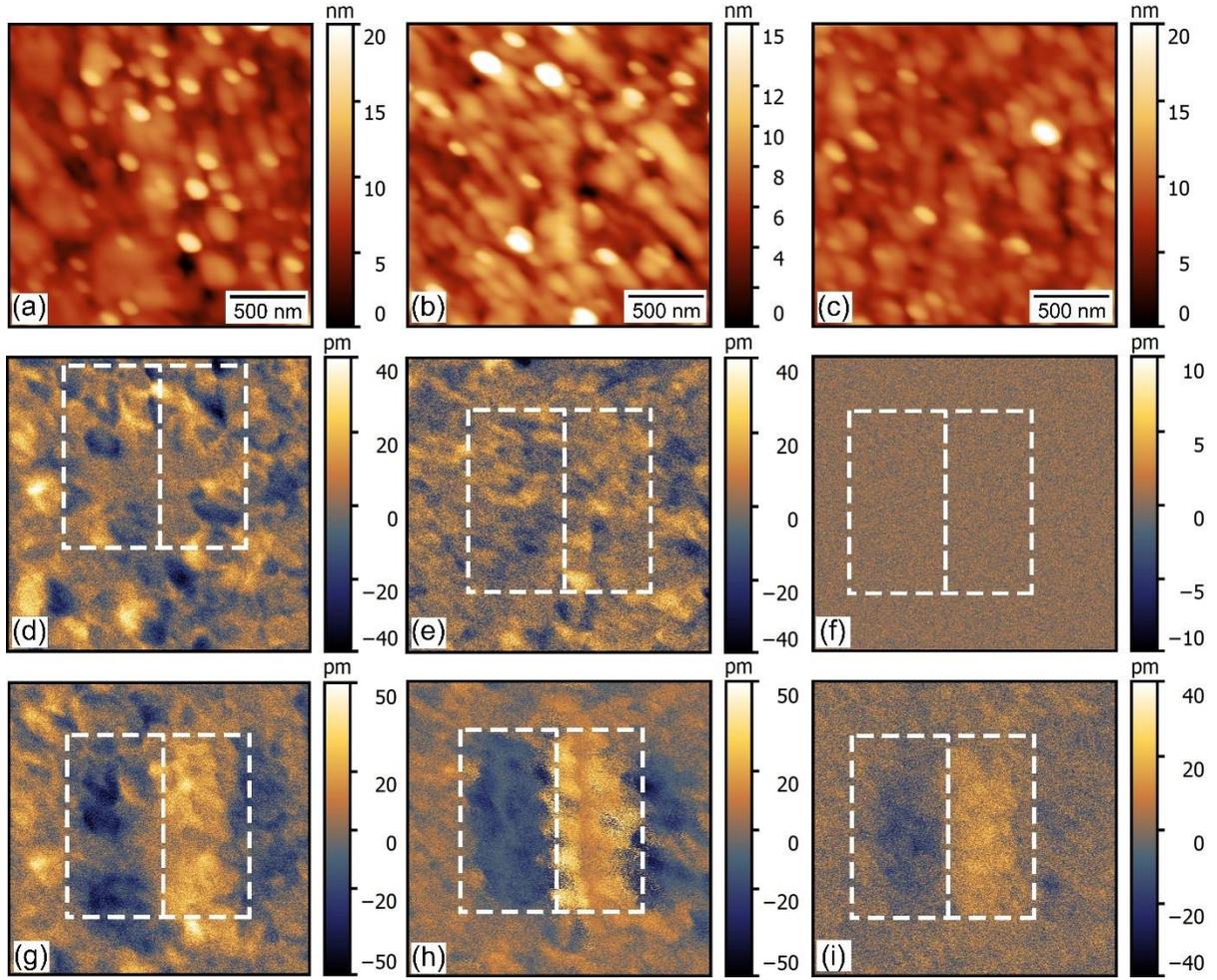

**Fig. 7.** In-plane PFM images. (a)-(c) Topography, RCosθ piezoresponse signal: (d)-(f) before and (g)-(h) after local poling of bi-square area by ±35 V DC voltage (left part is poled negatively, while right part - positively). (a), (d), (g) $0.8BiFeO_3 - 0.2BaTiO_3$; (b), (e), (h) $0.7BiFeO_3 - 0.3BaTiO_3$; (c), (f), (i) $(0.6)BiFeO_3-(0.4)BaTiO_3$-solid solutions. Displayed topography is measured simultaneously with PFM before poling and thereby shifted slightly after poling due to thermal drift.